# About the Discriminant Power of the Subgraph Centrality and Other Centrality Measures

# (Working paper)

Ernesto Estrada

Department of Mathematics and Statistics, University of Strathclyde, Glasgow G1 1XH, UK

The discriminant power of centrality indices for the degree, eigenvector, closeness, betweenness and subgraph centrality is analyzed. It is defined by the number of graphs for which the standard deviation of the centrality of its nodes is zero. On the basis of empirical analysis it is concluded that the subgraph centrality displays better discriminant power than the rest of centralities. We also propose some new conjectures about the types of graphs for which the subgraph centrality does not discriminate among nonequivalent nodes.

**Motivation and Definitions**

The use of graphs to represent complex systems in a variety of fields is nowadays a very established practice. In characterizing the importance of a node in a 'complex network' it is common to use the so-called *centrality measures* [1]. These centrality measures try to capture the ability of a node to communicate directly with other nodes, or its closeness to many other nodes or the quantity of pairs of nodes which need a specific node as intermediary in their communications determine many of the structural and functional properties of this node in a network.

Hereafter, we will consider only simple connected graphs having $n$ nodes and with adjacency matrices represented by $\mathbf{A}$. The simplest of all centrality measures is the node degree, which can be written as

$$k_i = (\mathbf{A}\mathbf{1})_i, \tag{1}$$

where $\mathbf{1}$ is an all-ones column vector. The degree of a node counts the number of nearest neighbours that a node has.

A different type of centrality tries to capture how close a node is to the rest of the nodes in a graph. It is known as the *closeness centrality* [1, 2], and can be expressed mathematically as follows

$$CC(u) = \frac{n-1}{s(u)}, \tag{2}$$

where

$$s(u) = \sum_{v \in V(G)} d(u,v), \tag{3}$$

is the sum of all shortest path distances $d(u,v)$ between the node $u$ and any other node in the graph.

A third centrality index was defined by considering that the information going from one node to another travels only through the shortest paths connecting those nodes. Then, if $\rho(i,j)$ is the number of shortest paths from node $i$ to node $j$, and $\rho(i,k,j)$ is the number of these shortest paths that pass through node $k$ in the network, the betweenness centrality of node $k$ is given by [1, 2]:

$$BC(k) = \sum_i \sum_j \frac{\rho(i,k,j)}{\rho(i,j)}, \quad i \neq j \neq k. \tag{4}$$

Obviously, this index characterizes the importance of a given node as intermediary in the communication among other nodes in the graph.

An extension of the degree centrality of a node was carried out by Bonacich by considering the principal eigenvectors of the adjacency matrix of a graph [1, 3, 4]. That is, let $\lambda_1$ be the largest eigenvalue of the adjacency matrix and let $\varphi_1$ be the eigenvector associated with it. Then, the eigenvector centrality of node $i$ is the corresponding entry of the principal eigenvector of the adjacency matrix: $\varphi_1(1)$. A nice interpretation of the eigenvector centrality can be obtained by using the following result [5]. Let us consider the number $N_k(i)$ of walks of length $k$ starting at node $i$ of a non-bipartite connected network. A walk of length $l$ is any sequence of (not necessarily different) nodes $v_1, v_2, \ldots, v_l, v_{l+1}$ such that for each $i = 1, 2, \ldots, l$ there is link from $v_i$ to $v_{i+1}$. A closed walk of length $l$ is a walk $v_1, v_2, \ldots, v_l, v_{l+1}$ in which $v_{l+1} = v_1$. Let

$$s_k(i) = N_k(i) \cdot \left[ \sum_{j=1}^{n} N_k(j) \right]^{-1}.$$ Then, for $k \to \infty$, the vector $[s_k(1), s_k(2), \cdots s_k(n)]^T$ tends towards the eigenvector centrality. It means that the eigenvector centrality of node $i$ can be seen as the ratio of the number of walks of length $k$ that departs from $i$ to the total number of walks of length $k$ in a non-bipartite connected network when the length of these walks is sufficiently large.

In 2005 Estrada and Rodríguez-Velázquez [1, 6] introduced the so-called subgraph centrality of a node in a graph. Because the number of closed walks of length $l$ starting (and ending) at a given node of a graph is given by $(\mathbf{A}^l)_{ii}$ we consider an index that penalizes the longest closed walks with respect to the shortest ones. That is,

$$EE(i) = \left( \sum_{l=0}^{\infty} \frac{\mathbf{A}^l}{l!} \right)_{ii} = (e^{\mathbf{A}})_{ii}, \tag{5}$$

which can also be written as

$$EE(i) = \sum_{j=1}^{n} \left[ \varphi_j(i) \right]^2 e^{\lambda_j}. \tag{6}$$

Then, the index measures the weighted participation of a node in all the subgraphs of a graph in a way that smallest subgraphs receive more 'importance' than the larger ones.

Because the main function of centrality indices is that of 'ranking' nodes according to a given topological property of the graph, it is important to know how well do they discriminate among the nonequivalent nodes of a graph. This is the question with which the current open problem deals.

**Discriminant power of centralities**

Let $\mathbf{c} = \begin{bmatrix} c_1 & c_2 & \cdots & c_n \end{bmatrix}^T$ be a vector of the centralities of the nodes in a graph, i.e., $c_i$ could be the degree, closeness, betweenness, eigenvector or subgraph centrality as described before. Let $s_c(G)$ be the standard deviation of the values of $\mathbf{c}$ for a graph $G$. We say that the nodes of the graph $G$ are not distinguishable by $\mathbf{c}$ if and only if $s_c(G) = 0$. The discriminant power $D(C)$ of a given index $C$ is then defined as the ratio of the number of graphs for which $s_c(G) = 0$ to the total number of graphs analyzed. The best way to perform this analysis is to consider those graphs that have the same number of nodes and reporting only the number of graphs for which $s_c(G) = 0$. Here we calculate $s_c(G)$ for all connected graphs having 5, 6, 7 and 8 nodes. The total number of graphs studied is 12,103. The results are resumed in the Table 1.

Table 1. Number of graphs with given number of nodes $n$ for which $s_c(G) = 0$ for the different centrality measures studied here.

| $n$ | subgraph | degree | eigenvector | closeness | Betweenness |
|---|---|---|---|---|---|
| 5 | 2 | 2 | 2 | 2 | 2 |
| 6 | 6 | 6 | 6 | 6 | 7 |
| 7 | 3 | 4 | 4 | 4 | 3 |
| 8 | 10 | 17 | 17 | 15 | 12 |

For the graphs with 5 nodes $s_c(C_5) = s_c(K_5) = 0$ for any **c**. For the graphs with 6 nodes there are five graphs for which $s_{SC}(G) = 0$. They are the cycle, complete graph, octahedral graph, utility graph and 3-prism graph. All these graphs are walk-regular [7]. A graph is walk-regular if, for each $l$, the number of closed walks of length $l$ starting a given node is independent of the selection of that node. It is know that vertex transitive and distance regular graphs are also walk-regular [7]. All the graphs previously mentioned are vertex transitive and in addition, the utility graph is also distance regular. All the other centrality measures do not distinguish the nodes in these graphs. In addition there is one more graph whose nodes are not distinguished by the betweenness centrality. The graph is illustrated in the Figure 1.

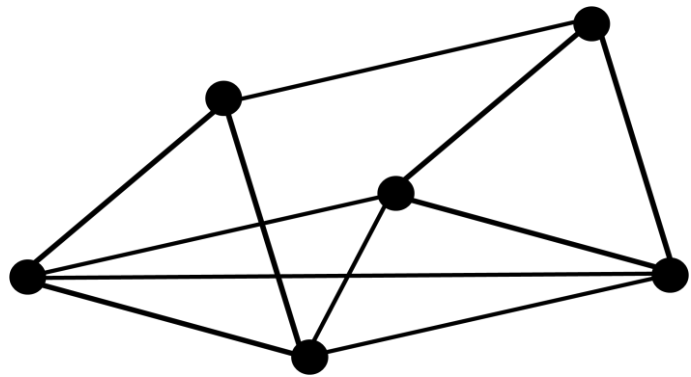

**Figure 1**. The graph having 6 nodes whose nodes are distinguishable by all centrality measures but by the betweenness centrality.

For the graphs having 7 nodes there are only three graphs which are not distinguishable by the subgraph centrality. They are the cycle, the complete graph and the 7-circulant graph (1,2). All of them are walk regular graphs. Neither of the other centralities distinguishes the nodes of these graphs. In addition, neither the degree nor the eigenvector centrality distinguishes between the nodes of the only regular graph with 7 nodes which is not walk regular.

The situation becomes more interesting for the graphs with 8 nodes. Here the subgraph centrality does not distinguish among the nodes of 10 graphs. These are the 10 walk regular graphs with 8 nodes. All these graphs have nodes which are not distinguishable by neither of the other centralities. The degree and eigenvector centralities do not distinguish the nodes of 17 graphs, i.e., all the regular ones, 10 walk-regular and 7 other regular ones, the closeness centrality does not discriminate the nodes of 15 graphs, i.e., the 10 walk-regular ones plus 5 other regular graphs, and the betweenness centrality is identical for all nodes in 12 graphs, i.e., the 10 walk-regular ones plus two graphs which are not regular. All these graphs are displayed in the Figure 2.

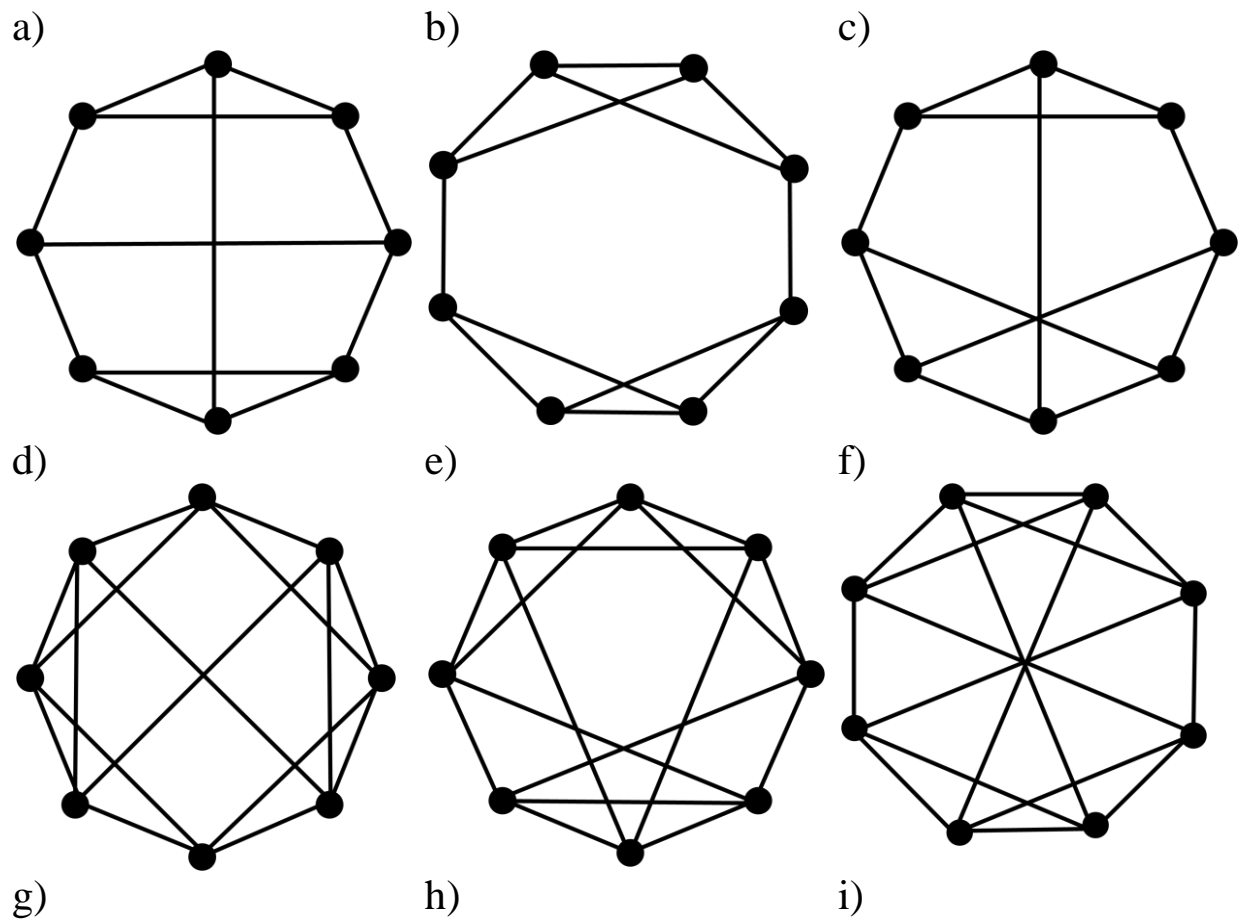

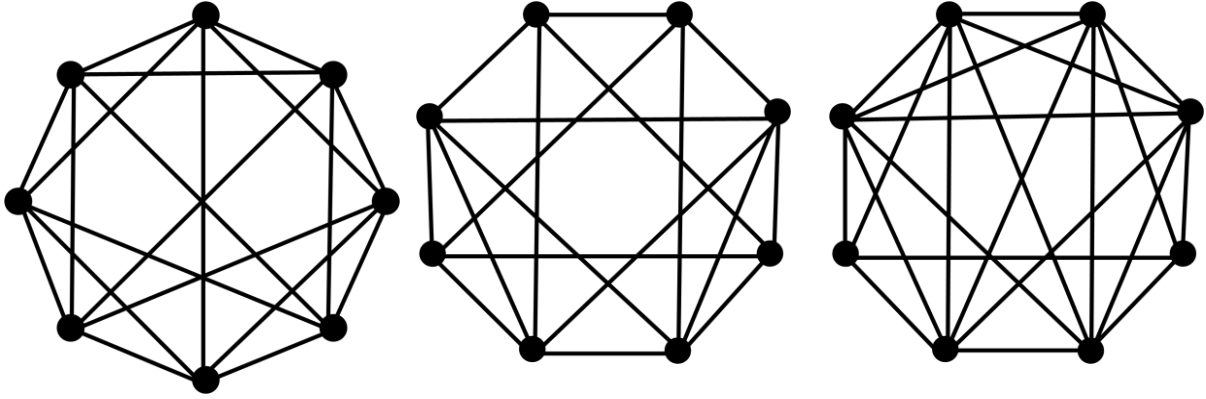

**Figure 2**. Examples of graphs with 8 nodes for which the discriminant power of the subgraph centrality if bigger that for the other centralities. The nodes of the graphs a)-g) are not differentiated neither by the degree nor the eigenvector centrality as they are regular graphs. The nodes of the graphs c)-g) are not distinguished by the closeness centrality, and the nodes in the graphs h) and i) are not differentiated by the betweenness centrality. In contrast, the subgraph centrality differentiates nonequivalent nodes in all these graphs.

**Old conjectures, counterexamples and new conjectures**

When the subgraph centrality index was proposed in 2005, Estrada and Rodríguez-Velázquez proposed the following conjecture.

**Conjecture 1** [6]. If $s_{sc}(G)=0$ then $s_c(G)=0$ for the degree, closeness, betweenness and eigenvector centrality.

That is, if there is a graph whose nodes are not distinguishable by the subgraph centrality, they are not distinguishable by the degree, closeness, betweenness and eigenvector centrality. Recently, this conjecture has been disproved by Puck Rombach and Porter [8]. They have found at least four graphs which are walk-regular but not distance-regular and for which $s_{sc}(G)=0$ but not for the closeness or betweenness centralities. On the basis of these results we propose the following stronger conjecture. Let $G_n$ be the set of all connected graphs with $n$ nodes. Let $H_n$ be the set of all walk-regular graphs with $n$ nodes which are not distance-regular. Let $R_n = G_n \setminus H_n$ be the set of all connected graphs with $n$ nodes which excludes those walk-regular, which are not distance-regular. Then,

**Conjecture 2**. $s_{sc} = 0 \Rightarrow s_c = 0$ for $C$ being the degree, closeness, betweenness and eigenvector centrality.

Based on the computational results we have shown previously we can also formulate the following conjecture.

**Conjecture 3**. $s_{sc}(G) = 0$ if and only if $G$ is walk regular.

The case that a walk regular graph has $s_{sc}(G) = 0$ is trivial from the definition of the subgraph centrality, but that the only graphs having nodes not discriminated by the subgraph centrality are the walk regular ones, is not trivial.

**References**

[1] E. Estrada, The Structure of Complex Networks, Theory and Applications; Oxford: Oxford University Press, 2011, Chapter 7.

[2] L. C. Freeman, Centrality in networks: I. conceptual clarification. Social Networks 1 (1979) 215–239.

[3] P. Bonacich, Factoring and weighting approaches to status scores and clique identification. J. Math. Sociol. 2 (1972) 113–120.

[4] P. Bonacich, Power and centrality: A family of measures. Am. J. Soc. 92 (1987) 1170-1182.

[5] D. Cvetković, P. Rowlinson, S. Simić, Eigenspaces of Graphs. Cambridge University Press, Cambridge, 1997.

[6] E. Estrada, J. A. Rodríguez-Velázquez, Subgraph centrality in complex networks. Phys. Rev. E 71 (2005) 056103.

[7] C. D. Godsil, B. D. McKay, Feasibility conditions for the existence of walk-regular graphs, Lin. Algebra Appl. 30 (1980) 51-61.

[8] M. Puck Rombach, M. Porter, Discriminating power of centrality measures. arXiv:1305.3146.